\documentclass[final,5p,times,twocolumn]{elsarticle}

\usepackage{graphicx}
\usepackage{amssymb}

\usepackage{lineno}

\usepackage[misc]{ifsym} 
\begin{document}

\begin{frontmatter}


\title{First-principles discovery of novel quantum physics and materials: \\ From theory to experiment}



\author[a1]{Yang Li}
\author[a1,b1,c1]{Yong Xu \Letter}
\ead{yongxu@mail.tsinghua.edu.cn}
\address[a1]{State Key Laboratory of Low Dimensional Quantum Physics and Department of Physics, Tsinghua University, Beijing, 100084, China}
\address[b1]{Frontier Science Center for Quantum Information, Beijing 100084, China}
\address[c1]{RIKEN Center for Emergent Matter Science (CEMS), Wako, Saitama 351-0198, Japan}

\begin{abstract}
Modern material science has been revolutionized by the discovery of novel topological states of quantum matter, which sheds new lights on solving long-standing scientific challenges. However, the exotic quantum phenomena are typically observable only in rare material systems under extreme experimental conditions. The search of suitable candidate materials that are able to work at ambient conditions is thus of crucial importance to both fundamental research and practical applications. Here we review our recent efforts on first-principles exploration of novel quantum physics and materials, focusing on   emergent quantum phenomena induced by spin-orbit coupling and its interplay with magnetism, topology and superconductivity. The first-principles material design guided by fundamental theory enables the discoveries of several key quantum materials, including next-generation magnetic topological insulators, high-temperature quantum anomalous Hall and quantum spin Hall insulators, and unconventional superconductors. A close collaboration with experiment not only successfully confirmed most of our theoretical predictions, but also led to surprising findings for further investigations, which greatly promotes development of the research field. 
\end{abstract}


\end{frontmatter}


\section{Introduction}

A few research subjects are well known to be long-standing grand challenges in condensed matter physics and material science, including low-power electronics, high-performance thermoelectrics and high-efficiency photovoltaics, which are of key importance to overcome the global energy and environment crisis. To make substantial advances, one has to go beyond traditional theoretical framework and tries to explore new physics and materials. Remarkably, revolutionary breakthroughs have been achieved in the emerging field of topological quantum materials during the last four decades~\cite{hasan2010,qi2011,haldane2017}. The findings of the integer quantum Hall effect, the fractional quantum Hall effect, as well as topological phase transitions and topological phases of matter were awarded the Nobel Prizes in Physics respectively in 1985, 1998 and 2016. Recent important progresses  include the experimental discoveries of the quantum spin Hall (QSH) effect~\cite{konig2007} and the quantum anomalous Hall (QAH) effect~\cite{chang2013}. These fundamental breakthroughs shed new lights on the challenging topics.

The use of ``topological concepts'' could indeed facilitate overcoming the challenges. For examples, the QSH and QAH insulators as new states of quantum matter provide topologically protected edge conduction channels that are immune from scattering, advantageous for low-dissipation devices and enhanced thermoelectric performance~\cite{hasan2010,qi2011,xu2014}. The QAH states or the surface states of topological insulators (TIs), when interacting  with superconductivity, will generate topological superconducting states with Majorana fermions that are antiparticles of their own and obey non-Abelian statistics, promising for fault-tolerant quantum computing~\cite{fu2008, qi2010prb,lian2018}. Moreover, topological nonlinear optical effects (e.g. shift current) are revealed to be originated from the geometric nature of Bloch wave functions and irrelevant to dissipative processes~\cite{morimoto2016,tokura2018}, which are useful for optoelectrics.

A key physical concept developed by modern condensed matter physics is the emergent electromagnetism~\cite{nagaosa2012}. The electromagnetic field is the most important and fundamental field in condensed matter. In addition to usual electromagnetic fields, several electromagnetic-like fields or gauge fields emerge in solids due to electron-electron and electron-ion interactions, such as the momentum-dependent magnetic field induced by the spin-orbit coupling (SOC), the magnetic field in the momentum space caused by Berry curvature, and the magnetic field in the real space introduced by spontaneous magnetization. These electromagnetic fields point new directions for the material research.

In this paper, we will review our recent efforts on first-principles exploration of novel quantum physics and materials for fundamental research and practical application, mainly focusing on two-dimensional (2D) materials with strong SOC effects. Whereas the SOC is usually neglected or treated as a perturbation in most previous works, the SOC effects have been recognized to play indispensable roles in creating many novel quantum phases, such as quantum materials with nontrivial topology, low-dimensional magnetism stabilized by magnetic anisotropy, and unconventional superconductivity (e.g., topological superconductivity, Ising superconductivity). Thus it is a fertile field to search for emerging new physics and materials by coupling magnetism, topology and superconductivity with strong SOC. Moreover, the study of 2D materials, which is interested by the prominent quantum effects and rich material properties but demands atomic-resolution knowledge to fully understand and control, offers new opportunities and challenges for the material research. 
The 2D systems are simple and interesting in theory, but complex and somewhat uncontrollable in experiment. The gap between theory and experiment can be bridged by first-principles calculations and material design.

In the following, we will first review the first-principle discovery of intrinsic magnetic TIs (MnBi$_2$Te$_4$- and LiFeSe-family materials) and our continuing efforts to pursue high-temperature QAH effects. Then we will present the prediction of a new family of 2D materials (stanene and its derivatives) and show the novel physics discovered in the materials, including large-gap QSH states, unusual defect physics, topological thermoelectric size effects, and unconventional superconductivity. Finally we will give a brief summary and an outlook for future research.

\section{Discovery of intrinsic magnetic TIs MnBi$_2$Te$_4$ and LiFeSe: towards high-temperature QAH effects}

TIs are new states of quantum matter characterized by a nontrivial $\mathbb {Z} _{2}$ topology and the existence of spin-momentum locked gapless boundary states within the insulating bulk gap~\cite{hasan2010,qi2011}. The gapless feature of topological boundary states is protected by time reversal symmetry (TRS). The well-known TIs thus are also called time-reversal invariant (TRI) TIs. Once the TRS is broken by spontaneous magnetization, the topological boundary states would generally become gapped, and a new class of topological quantum materials, magnetic TIs, can be generated. Intensive research effort has been devoted to study magnetic TIs, in which the interplay of topology and magnetism offers new opportunities to explore emergent quantum physics (e.g., the QAH effect, axion electrodynamics, Majorana fermions) and potential applications (e.g., low-power electronics and quantum computing)~\cite{tokura2019}.

A key material is the QAH insulator, which is a magnetic TI in 2D and could be used as build blocks to create 3D magnetic topological materials. The QAH insulators, also named Chern insulators, distinguish from normal 2D insulators by a nonzero integer Chern number ($C$), a quantized anomalous Hall conductance ($\sigma^{A}_{xy} = C e^2/h$), and the appearance of linear dispersive chiral states on the edges~\cite{he2014,weng2015,liu2016}. They are able to give the quantum-Hall like effects without the aid of external magnetic field, attractive for device applications.

\begin{figure}
  \includegraphics[width=0.49\textwidth]{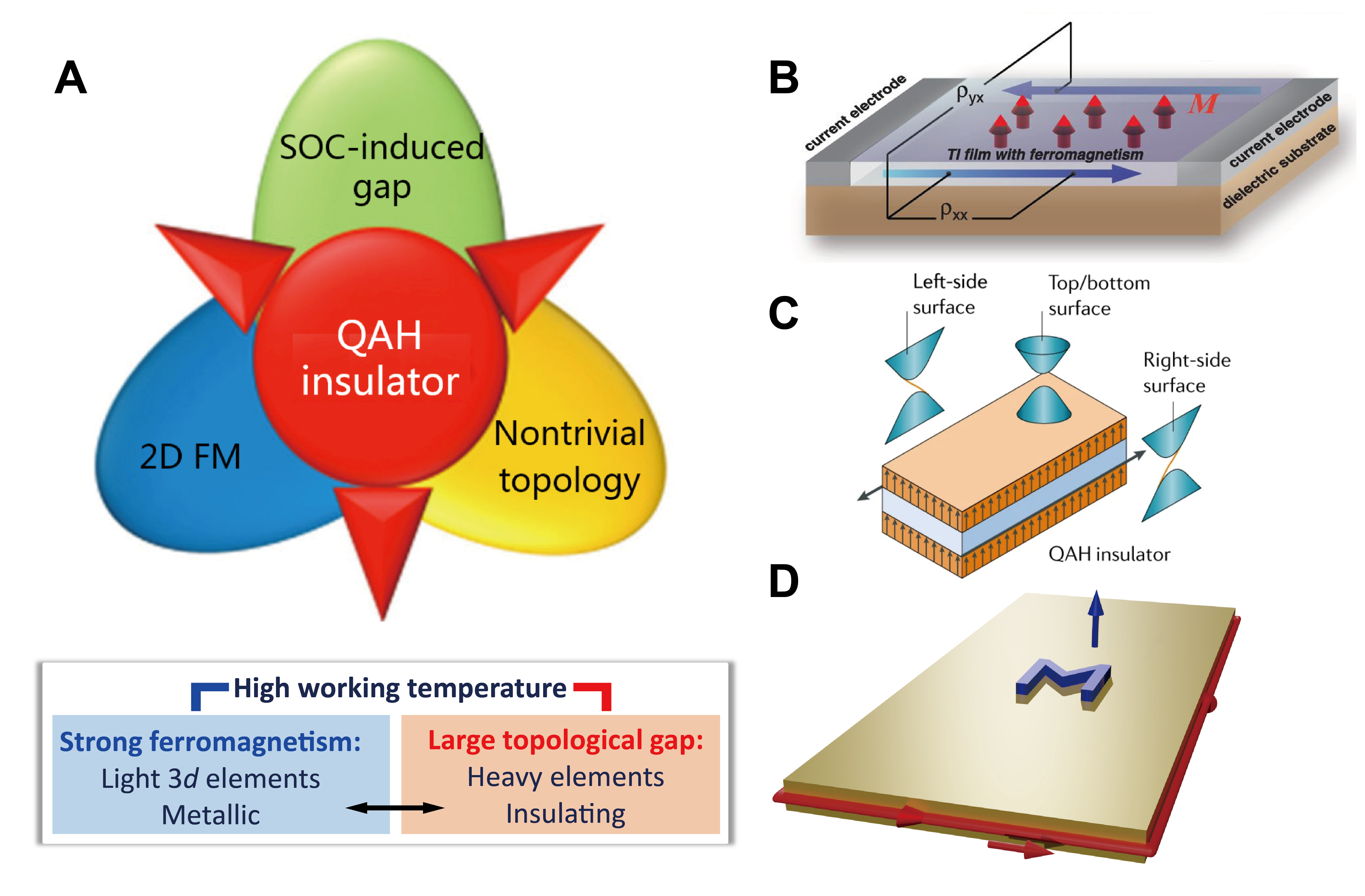}
  \label{fig1}
  \caption{(A) Physical requirements of QAH insulators, which should simultaneously have a 2D FM order, an insulating bulk gap opened by the SOC, and a nontrivial band topology. The high working temperature requires both strong 2D ferromagnetism and large topological band gap, which have conflicting material requirements. (B-D) Different kinds of magnetic topological materials: (B) magnetically doped into TIs, (C) magnetic-topological heterostructures, and (D) intrinsic magnetic TIs. Figures (B) from ~\cite{chang2013} and (C) from ~\cite{tokura2019}.}
\end{figure}

The QAH effect was first theoretically proposed by Haldane in his seminal work, where a quantum Hall effect without Landau levels is predicted by adding a periodic zero-flux magnetic field into a honeycomb-net model~\cite{haldane1988}. In 2013, such kind of effect was first experimentally discovered in magnetically doped TI films~\cite{chang2013}, opening a new era of quantum material research. In the early experiments, complicated quaternary alloys Cr- or V-doped (Bi$_{x}$Sb${_{2-x}}$)Te$_{3}$ were used and their working temperatures are ultralow, on the order of tens of miniKelvins~\cite{chang2013,checkelsky2014,kou2014,chang2015}. For many years, they kept as the only materials available for experiment, and their working temperature has been enhanced to $\sim$2 K by a magnetic modulation doping technique~\cite{mogi2015}, but can hardly be further improved. In fact, very few research groups could complete the experiment due to the complicated material growth and the extreme experimental conditions. This forbids in-depth experimental studies and practical device applications. The bottleneck problem is how to find better QAH materials, preferably with simpler atomic structures and higher working temperatures.

\subsection{Guiding principles for seeking QAH materials}

Let us first discuss the physical requirements of QAH insulators~\cite{he2014,weng2015,liu2016}. The topological Chern number is defined as an integral of Berry curvature over the 2D momentum space. For TRI systems, the Berry curvature is an odd function of the momentum, implying a vanishing $C$. It is thus necessary to break TRS for obtaining a nonzero $C$. For that purpose, one may apply magnetic or Coriolis fields or introduce a spontaneous magnetization into materials. The latter way that preserves the lattice translational symmetry and enables the use of Bloch theorem is relevant for the discussion of QAH insulators. Physically, the nonzero $\sigma^{A}_{xy}$ is generated either by a non-vanishing magnetization in combination with SOC or by a non-collinear antiferromagnetic (AFM) structure with a scalar spin chirality~\cite{nagaosa2010,nagaosa2012,zhou2016,feng2020}. For simple magnetic materials that typically have collinear spins and include one type of magnetic atoms, the resulting QAH insulators should simultaneously have a 2D ferromagnetic (FM) order, an insulating bulk gap opened by the SOC, and a nontrivial band topology.

Finding high-temperature FM semiconductor is a well-known challenging task~\cite{ando2006}. For comparison, seeking high-temperature QAH insulators is even more challenging (illustrated in Fig. 1A), since the QAH insulators belong to a special kind of FM semiconductors satisfying much more strict material requirements: they are crystallized in 2D and have topological electronic structure with band gap induced by SOC. One fundamental problem invoked by the Mermin-Wagner theorem is that the FM order in 2D is easily destroyed by thermal fluctuations and maybe not experimentally realizable~\cite{mermin1966}. Recent works found that the 2D ferromagnetism can be stabilized by magnetic anisotropy and survive at finite temperatures~\cite{gong2019science}. A few 2D FM semiconductors have been obtained experimentally mainly from van der Waals (vdW) layered materials, such as CrI$_{3}$ and Cr$_{2}$Ge$_{2}$Te$_{6}$~\cite{huang2017,gong2017}. Nevertheless, they all have low Curie temperatures (below the nitrogen temperature), and more importantly none of them are topologically nontrivial.

From the material point view, a high-temperature QAH insulator should meet both criteria of strong ferromagnetism and large SOC-induced topological gap (Fig. 1A). While metallic systems composed of light $3d$ elements are preferred by the former criterion, insulating materials comprised by heavy elements are demanded by the latter~\cite{li2020prl}. The conflicting material requirements make the pursuit challenging. Previous works have tried to construct magnetic topological materials via two compromising ways: introducing magnetism extrinsically into TIs either by applying magnetic doping or alloying (Fig. 1B) or by fabricating magnetic-topological heterostructures (Fig. 1C). The major drawback of the resulting material systems is that the extrinsic magnetic effects are usually weak, easily affected by structural disorders or defects, and very difficult to control in experiment. Although the QAH effect was originally realized in the first class of materials, one can hardly improve the working temperature limited by the strong disorder effects~\cite{tokura2019}. Many groups have attempted to find the QAH states in the second class of materials, but no successful results was reported. One promising direction is to find the next-generation QAH materials---intrinsic magnetic TIs (Fig. 1D), which are stoichiometric magnetic compounds inherently possessing both intrinsic magnetism and topological electronic structures~\cite{li2019sci_adv,gong2019}. However, despite long-time systematic experimental search, no vdW layered materials has been found to fit the requirements of intrinsic magnetic TIs (before the discovery of candidate material MnBi$_2$Te$_4$ in 2019). In this context, first-principles material predictions become indispensable for the development of this field.

\begin{figure}
  \includegraphics[width=0.49\textwidth]{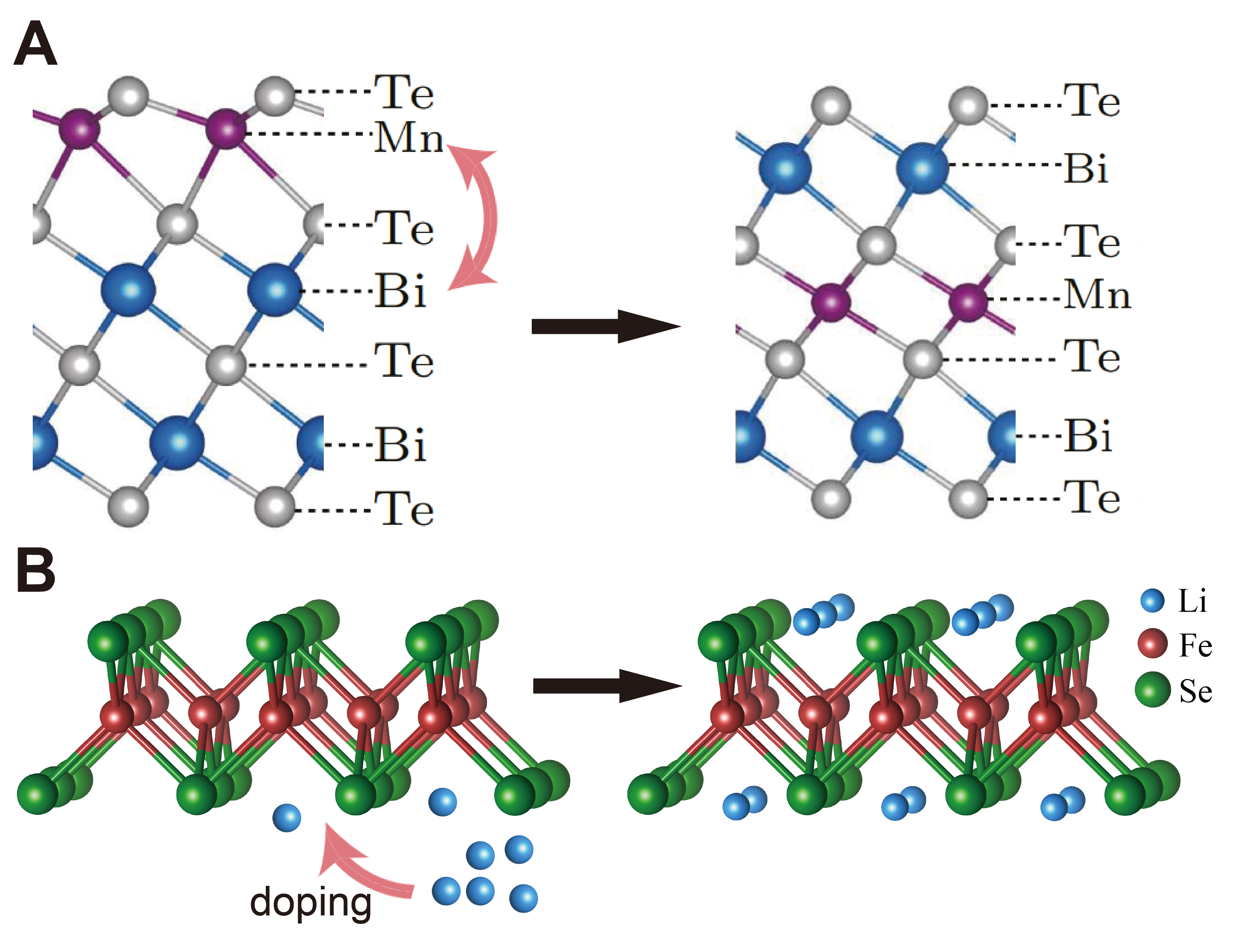}
  \label{fig2}
  \caption{Two different strategies to construct intrinsic magnetic TIs from existing vdW layered materials: (A) magnetic layer intercalation and (B) Li decoration. Figures (A) adapted from ~\cite{gong2019} and (B) adapted from ~\cite{li2020prl}.}
\end{figure}

In the following, we will present two strategies to transform the existing vdW layered materials into intrinsic magnetic TIs (including QAH insulators) as illustrated in Fig. 2. Advantages of vdW material systems (e.g., easy to prepare thin-film samples and construct heterostructures) thus could be inherited by magnetic TIs. The first strategy is to intercalate a magnetic layer into a vdW layered TI, namely magnetic layer intercalation (Fig. 2A)~\cite{li2019sci_adv,gong2019}. For instance, the intercalation of a bilayer Mn-Te into each quintuple-layer Bi$_2$Te$_3$ creates a vdW-coupled septuple-layer compound MnBi$_2$Te$_4$. MnBi$_2$Te$_4$ is an intrinsic magnetic TI that beautifully combines the ferromagnetism induced by Mn and the nontrivial band topology contributed by Bi-Te. The other strategy is to decorate Li into vdW layered materials containing magnetic elements (Fig. 2B). The Li decoration alters the electron occupation of magnetic elements, which might dramatically change the magnetic and topological properties of the mother material. As an example, we will demonstrate that a monolayer LiFeSe, created by decorating Li on the superconducting material FeSe monolayer, is chemically and physically distinct from FeSe. The new material is a large-gap QAH insulator with extremely robust 2D ferromagnetism, which could serve as a promising material candidate to realize high-temperature QAH effects~\cite{li2020prl}.

\subsection{Discovery of intrinsic magnetic TI MnBi$_2$Te$_4$}

\begin{figure}
  \includegraphics[width=0.49\textwidth]{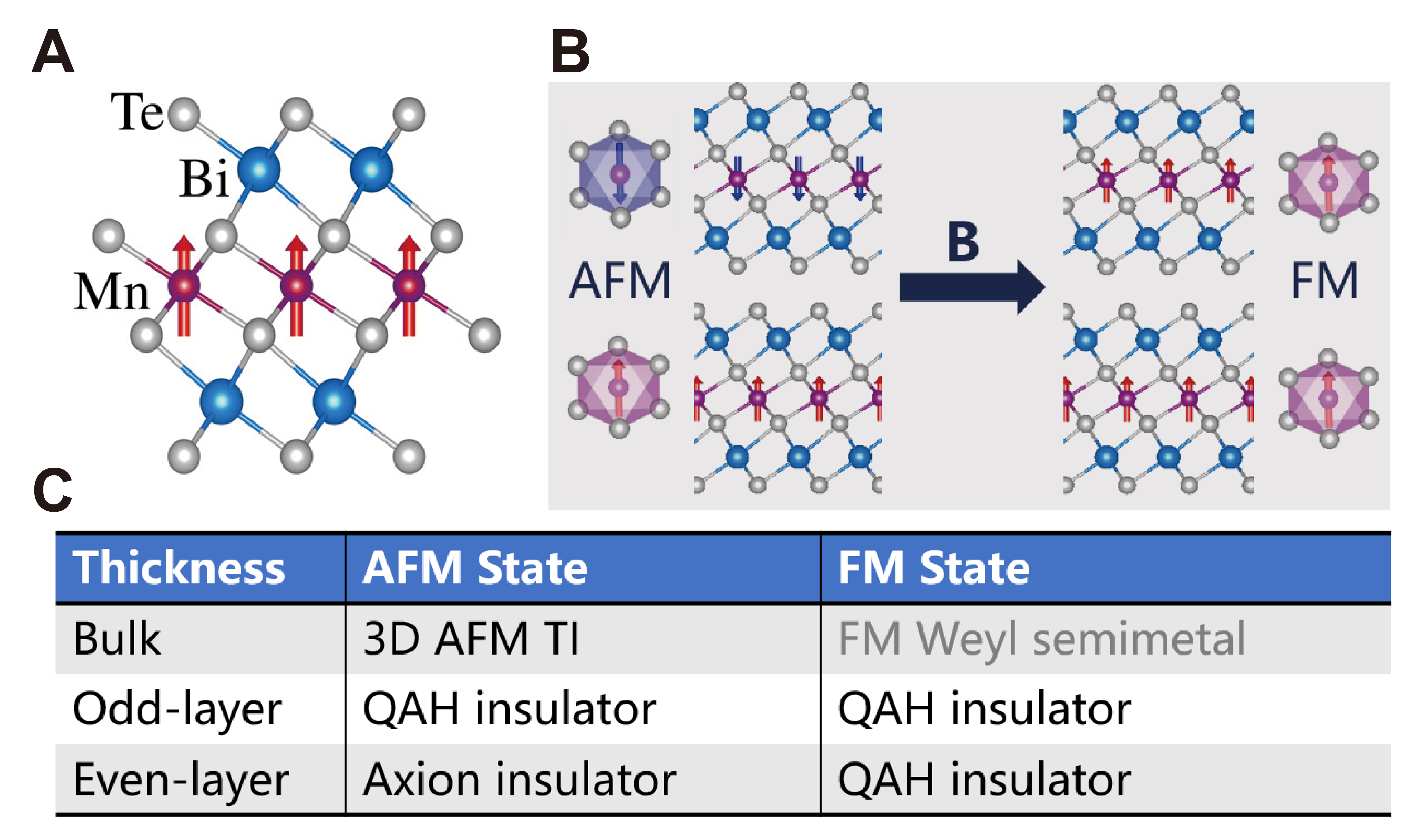}
  \label{fig3}
  \caption{(A,B) Atomic structures and magnetic configurations of MnBi\(_2\)Te\(_4\) (A) monolayer and (B) bulk. The A-type AFM ground state can be transformed into the FM state by applying external magnetic field {\bf{B}}. (C) Rich topological properties of MnBi\(_2\)Te\(_4\) with varying thickness/dimensions in different magnetic states. Figures (A) from ~\cite{li2019sci_adv} and (B) adapted from ~\cite{gong2019}.}
\end{figure}

Previous studies of QAH states predominately focused on magnetically doped TIs. In contrast, we find that the magnetic layer intercalation can generate a series of intrinsic magnetic TIs---the MnBi$_2$Te$_4$-family materials, which provide a useful material platform to study magnetic topological physics, including the QAH effect, topological magnetoelectric effects and topological superconductivity~\cite{li2019sci_adv}.

\subsubsection{Atomic, electronic and magnetic structures of MnBi$_2$Te$_4$}
The concept of magnetic layer intercalation is well illustrated by the epitaxial growth of MnBi$_2$Te$_4$~\cite{li2019sci_adv,gong2019}. For a Te-Mn bilayer adsorbed on a Te-Bi-Te-Bi-Te (Bi$_2$Te$_3$) quintuple, an atom swapping between Mn and Bi lowers the energy by $\sim$0.5 eV/unit cell, leading to a stable septuple-layer (SL) Te-Bi-Te-Mn-Te-Bi-Te (MnBi$_2$Te$_4$). Such kind of process can be realized by post-annealing, as demonstrated by molecular beam epitaxy (MBE) growth and confirmed by atomic-resolution scanning transmission electron microscopy~\cite{gong2019}. Magnetic layers of Mn are thus intercalated into the TI material Bi$_2$Te$_3$ (Fig. 2A). Similarly, other members of the MnBi$_2$Te$_4$-family materials, including $MB_2T_4$ ($M$ = V, Mn, Ni, or Eu, $B$ = Bi or Sb, $T$ = Te, Se, or S) are predicted to be stable~\cite{li2019sci_adv}.

MnBi$_2$Te$_4$ has a rhombohedral layered structure (space group $R\bar{3}m$), in which the neighboring SLs are coupled by vdW interactions (Fig. 3A,B). The intralayer (interlayer) magnetic coupling is FM (AFM), showing an A-type AFM configuration with an out-of-plane easy axis (Fig. 3B)~\cite{li2019sci_adv,gong2019,zhang2019prl,otrokov2019}. Single-layer MnBi$_2$Te$_4$ is a 2D FM semiconductor that has a topologically trivial band gap opened by the strong quantum confinement. A topological band inversion happens between the $p_z$ orbitals of Bi and Te in MnBi$_2$Te$_4$ bulk, induced by the strong SOC of Bi-Te. Therefore, MnBi$_2$Te$_4$ is an intrinsic magnetic TI with magnetic states from Mn and topological states from Bi-Te. 

\subsubsection{Prediction of rich topological physics in MnBi$_2$Te$_4$}

Remarkably, the unique vdW layered structure and A-type AFM order of MnBi$_2$Te$_4$ facilitate exploring the interplay of dimension, magnetism, and topology, which leads to rich topological states as summarized in Fig. 3D~\cite{li2019sci_adv}. Specifically, a transition from AFM to FM states can be driven by applying moderate magnetic fields ($\sim$5 T), benefiting from the weak interlayer exchange coupling and large magnetic moment (5 $\mu_B$) of Mn; a magnetic transition to paramagnetic (PM) state occurs when increasing the temperature above the N\'{e}el temperature ($T_N \sim$ 25 K)~\cite{gong2019,otrokov2019,cui2019,liu2020nat_mater}. Moreover, the sample thickness of vdW layered structures can be well controlled experimentally, which enables the tuning of quantum confinement effects and the change of physical dimension from 2D to 3D. The different magnetic states and physical dimensions generate a series of novel states of quantum matter in MnBi$_2$Te$_4$, including AFM TI, FM Weyl semimetal (WSM) and TRI TI in 3D as well as QAH, axion, and QSH insulators in 2D, as we will discuss below.

\begin{figure}
  \includegraphics[width=0.49\textwidth]{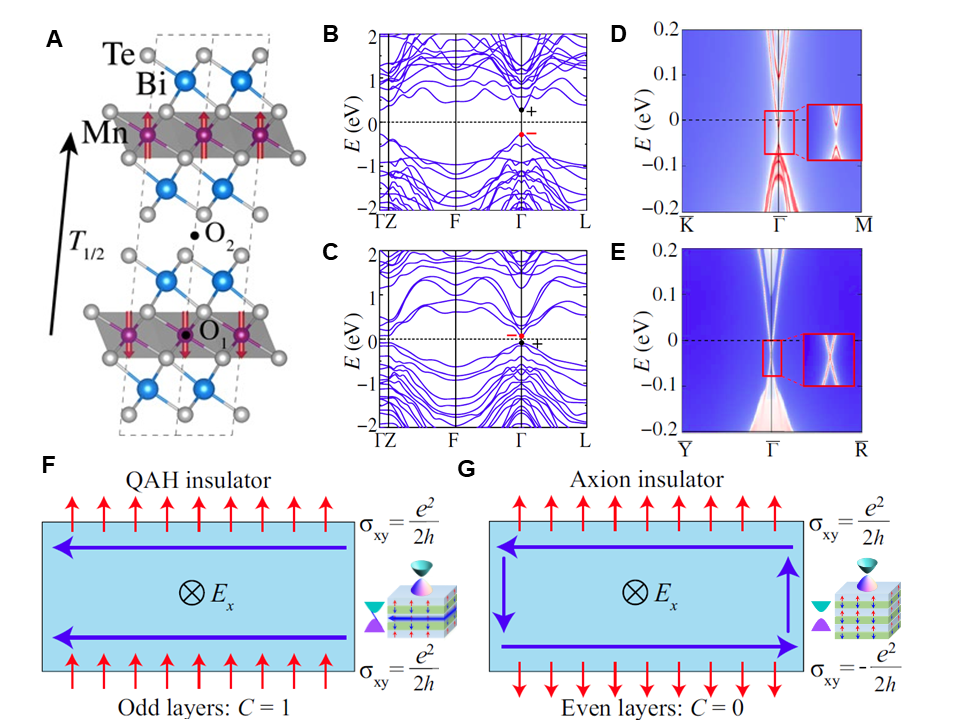}
  \label{fig4}
  \caption{(A) Atomic and magnetic structures, band structures (B) excluding and (C) including the SOC, (D) gapped (111) and (E) gapless (110) topological surface states of 3D MnBi\(_2\)Te\(_4\) in its magnetic ground state. (F,G) Oscillating change between (F) QAH and (G) axion insulators for thin films of AFM MnBi\(_2\)Te\(_4\). Figure adapted from ~\cite{li2019sci_adv}.}
\end{figure}

The bulk MnBi$_2$Te$_4$ in its magnetic ground state is a novel topological quantum matter---3D AFM TI (Fig. 4A-E), which is protected by the $S = \Theta T_{1/2}$ symmetry (the combination of time-reversal $\Theta$ and a primitive-lattice translation $T_{1/2}$), classified by a $\mathbb {Z} _{2}$ topological invariant, and featured by a quantized magnetoelectric effect~\cite{mong2010}. This magnetic topological state was theoretically proposed by Mong et al. in 2010, but has not been experimentally confirmed for a long time. In the original work, a model of AFM TI is constructed by staggering quantum Hall layers~\cite{mong2010}. In contrast, the AFM TI state is formed by the A-type AFM order and the SOC-induced band inversion in MnBi$_2$Te$_4$ (Fig. 4A-C)~\cite{li2019sci_adv,zhang2019prl}, whose material requirements are experimentally more friendly.

A hallmark feature of AFM TI is the existence of topological surface states which are gapped on the top/bottom (111) surfaces but remain gapless on the side surfaces (Fig. 4D-E). The topological surface Dirac fermions generally will acquire a mass term from spontaneous magnetization. However, the Dirac mass term vanishes on the side surfaces due to the preservation of $S$ symmetry. This is not the case on the top/bottom (111) surfaces, where the Dirac gap is as large as 40 to 50 meV at zero temperature~\cite{li2019sci_adv,gong2019}. The existence of this magnetization-induced gap is essential to realize topological magnetoelectric effects and explore axion electrodynamics~\cite{wilczek1987,qi2008,essin2009}.

Thin films of AFM MnBi$_2$Te$_4$ display an intriguing oscillating change between QAH and axion insulators, dependent on the parity of the number of layers (Fig. 4F,G)~\cite{li2019sci_adv}. The key physics is that the interaction of topological surface states with 2D ferromagnetism of the surface layer generates the so called half quantum Hall effect ($\sigma_{xy} = e^2/2h$)~\cite{qi2011}, which has opposite signs for opposing surface magnetizations. As a result, the two surface contributions are of the same sign and add up to a quantized Hall conductance $\sigma_{xy} = e^2/h$ in an odd-layer film, corresponding to the QAH state with $C=1$; whereas they compensate with each other in an even-layer film, giving rise to an axion insulator state with $C=0$. Distinct from trivial insulators, the axion insulator is characterized by a plateau of zero Hall conductance and can display quantized magnetoelectric or magneto-optical effects~\cite{morimoto2015,wang2015,wu2016,mogi2017,mogi2017_2,xiao2018}. As the Chern number varies from even-layer to odd-layer films, the gapless chiral channels would appear at their boundaries (i.e., along the step edges), which might be detected by scanning tunning microscopy (STM). Superior to previous proposals, the QAH and axion insulator states predicted in MnBi$_2$Te$_4$ are intrinsic, in no need of external magnetic effects.

\begin{figure}
  \includegraphics[width=0.49\textwidth]{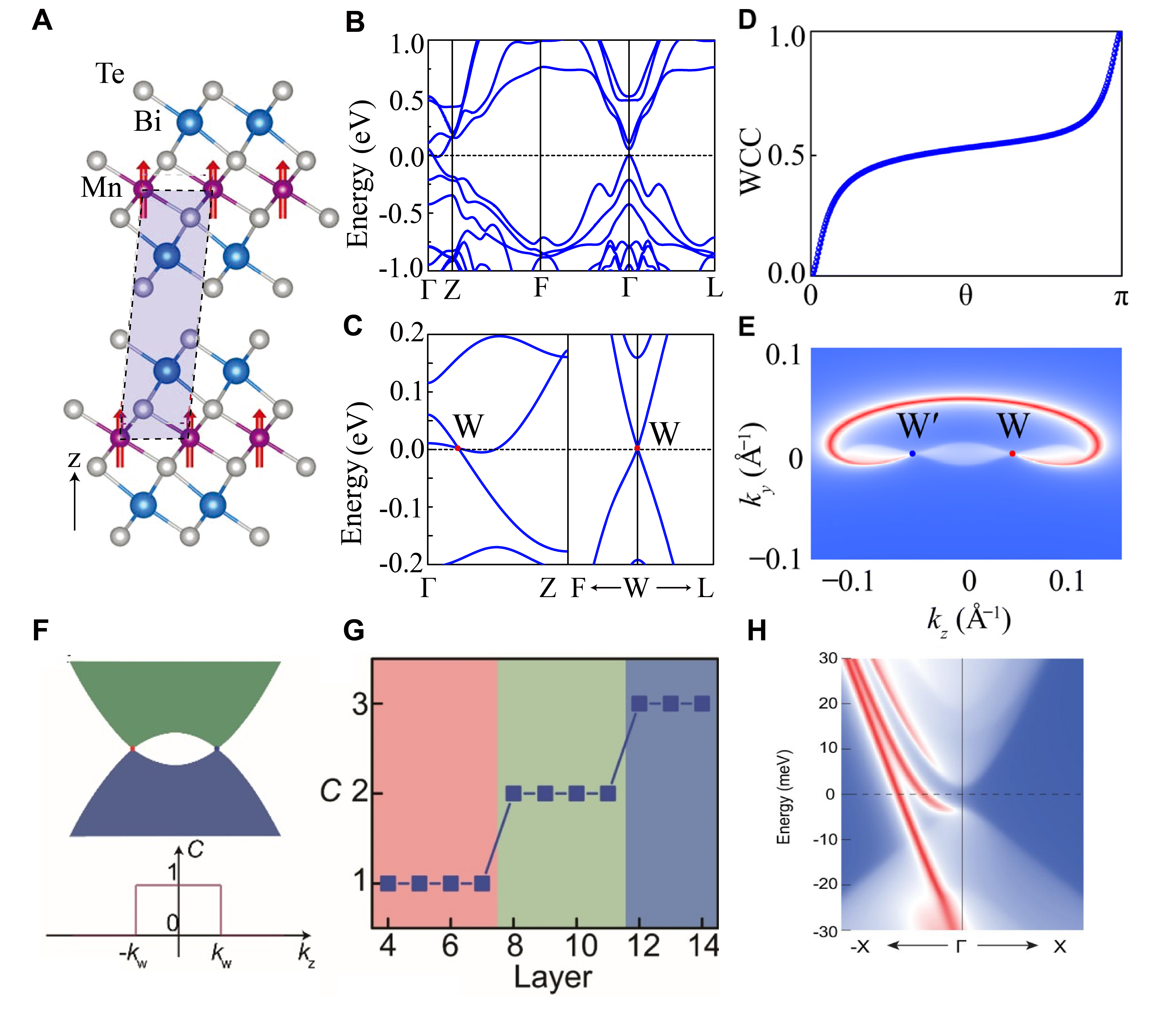}
  \label{fig5}
  \caption{(A) Atomic and magnetic structures, (B,C) band structures, (D) motion of Wannier charge center (WCC), and (E) Fermi-arcs surface states of the \((1\bar{1}0)\) surface of 3D MnBi\(_2\)Te\(_4\) in the FM state.  (F) A schematic diagram displaying the band structure along the \(k_z\) direction and the \(k_z\)-dependent Chern number for FM MnBi\(_2\)Te\(_4\). (G) The calculated Chern number as a function of film thickness. (H) Calculated topological edge states along the (100) direction for 9-SL FM MnBi\(_2\)Te\(_4\). Figures (A-E) from ~\cite{li2019sci_adv} and (F-H) from ~\cite{ge2020}.}
\end{figure}

The FM MnBi$_2$Te$_4$ bulk is a topological WSM (Fig. 5A-E) that hosts various emergent quantum physics, including Fermi arc surface states, the chiral anomaly, anomalous Hall effect and axion electrodynamics~\cite{armitage2018}. When the magnetic configuration of MnBi$_2$Te$_4$ is transformed from AFM to FM under magnetic fields, the spin degeneracy is destroyed, and the interlayer electronic coupling gets significantly enhanced due to the breaking of $PT$ symmetry (the combination of spatial inversion $P$ and time reversal $T$)~\cite{li2019prb}. This changes the topological band inversion along the $\Gamma$-$Z$ direction in the momentum space (Fig. 5C), leading to a quantum phase transition from AFM TI to FM WSM~\cite{li2019sci_adv,zhang2019prl}. The band-crossing points of WSM, namely Weyl points, are monopoles of momentum-space Berry curvature, which is characterized by nonzero quantized Berry fluxes $2\pi C$ ($C$ is the Chern number or topological charge), topologically protected against perturbations and can only be annihilated pairwise~\cite{armitage2018}. The TRI WSMs must be even pairs of Weyl points,  for instance, 12 pairs in TaAs~\cite{xu2015,lv2015}. Odd pairs of Weyl points are obtainable only in magnetic WSMs. Remarkably, FM MnBi$_2$Te$_4$ has one simple pair of Weyl points, representing the simplest WSM. Since the WSM states couple strongly with the magnetism in MnBi$_2$Te$_4$, their properties can be greatly tuned by varying the orientation of magnetization via magnetic fields~\cite{li2019prb}. The simple and tunable features ensure that MnBi$_2$Te$_4$ is a promising candidate material to study Weyl physics.

Intriguing QAH physics will emerge in thin films of FM MnBi$_2$Te$_4$ (Fig. 5F-H). Both Weyl points of the bulk are projected onto the $\Gamma$ point of the (111) surface Brillouin zone, and thus can couple with each other in the 2D flakes, opening insulating band gaps. Theoretical calculations indicate that multi-layer FM MnBi$_2$Te$_4$ are QAH insulators, whose Chern number increases abruptly with film thickness (Fig. 5G)~\cite{ge2020}. The thickness-dependent Chern insulator states in 2D is a manifest of the FM WSM state in 3D~\cite{xu2011,yokomizo2017}. This could be understood as follows. For FM MnBi$_2$Te$_4$, the $k_z$-dependent Chern number $C(k_z)$, defined for every 2D $k_x$-$k_y$ momentum space with a given $k_z$, is quantized to one in the band-inversion region near the $\Gamma$ point and decreases abruptly to zero once crossing the Weyl points (Fig. 5F). When changing from 3D bulk to 2D films, $k_z$ is allowed to take quantized values by quantum confinement. The more $k_z$ values located within the band inversion region, the larger the Chern number in total, implying a growth of Chern number with film thickness.

Temperature plays a critical role in determining the material properties of MnBi$_2$Te$_4$. When increasing the temperature above $T_N$, a magnetic transition from AFM to PM states happens, the TRS gets recovered, and the system changes from AFM TI to TRI TI, giving gapless topological Dirac fermions on all the surfaces~\cite{li2019sci_adv,gong2019}. In particular, the Dirac gap of the (111) surfaces would gradually decrease to zero when increasing the temperature above $T_N$. Such kind of temperature-dependent feature can serve as a smoking gun evidence of the magnetization-induced gap. Moreover, MnBi$_2$Te$_4$ thin films in the PM state are no longer QAH or axion insulators, but oscillate between the QSH and trivial insulators for varying film thickness, as predicted in Bi$_2$Te$_3$~\cite{liu2010prb}.

Another interesting issue is the interlayer magnetic coupling (IMC), which dictates the magnetic and topological properties of the system. First-principles calculations revealed that the IMC of MnBi$_2$Te$_4$-family materials is unusually long-range, caused by the superexchange coupling mediated by the delocalized $p_z$ orbitals across the vdW gap, and its sign (FM or AFM) is tunable by varying the $d$ orbital occupations of neighboring layers~\cite{li2020prb}. By tuning the IMC in the vdW heterostructures, the FM WSM and high-Chern-number QAH phases can be realized without applying external magnetic field. Importantly, a FM high-order TI phase is discovered in the material system, which displays gapless chiral hinge modes along both in-plane and out-of-plane directions, forming the so called ``pseudo-3D'' QAH insulator states~\cite{li2020prb,zhang2020prl,tanaka2020}. The real space distribution of ``pseudo-3D'' QAH insulator states can be controlled by manipulating the magnetic orientation, which is useful for designing ``topological magnetic switch''~\cite{zhang2020prl}.

In addition, theoretical works on MnBi$_2$Te$_4$ proposed many different means to control material properties, for instance, the creation of chiral Majorana modes by proximity coupling between AFM TI and $s$-wave superconductor~\cite{peng2019}, the stabilization of surface magnetism by exchange bias in MnBi$_2$Te$_4$/CrI$_3$ heterostructures~\cite{fu2020}, and the engineering of Berry curvature and topological phase transition by electric gating in MnBi$_2$Te$_4$ films~\cite{du2020}. Considering the fast growth of the field, a detailed review of all the relevant works are outside the scope of this paper.

\subsubsection{Recent experimental progresses on MnBi$_2$Te$_4$}
In this part, the most relevant experimental progresses will be briefly reviewed. The intrinsic magnetic TI states of MnBi$_2$Te$_4$ were first experimentally investigated by two research groups. Our experimental collaborators fabricated MnBi$_2$Te$_4$ films by MBE, detected an A-type AFM order by superconducting quantum interference device (SQUID) and Hall measurements, and observed gapless topological surface states by angle-resolved photoemission spectroscopy (ARPES) at temperatures near $T_N$, which for the first time confirms the existence of AFM TI states~\cite{gong2019}. Notice that the magnetization-induced gap was not detected in this early experiment, which cannot lower the temperature below $T_N$. The other group grew MnBi$_2$Te$_4$ single crystals, found similar magnetic order, and observed gapped topological surface states by ARPES at varying temperatures below and above $T_N$~\cite{otrokov2019}. However, the claimed topological Dirac gap does not close even far above $T_N$, excluding its correlation with magnetism. In contrast, high-resolution ARPES measurements of the following works indicate that the apparent Dirac gap is very likely contributed by bulk states, and a diminished surface Dirac gap is always observed, independent of temperature~\cite{hao2019prx,li2019prx,chen2019prx}. These data also contradict with theoretical results that expect a gradual opening of surface Dirac gap by lowering temperature. One possible reason is that the surface magnetic structures are disordered in the experimental samples that have many structural defects. In-depth studies are needed to understand the problem.

Recently impressive research progresses have been made on the magnetoelectric transport studies of MnBi$_2$Te$_4$. The intrinsic QAH effect has been successfully observed in an odd-layer (5 SLs) MnBi$_2$Te$_4$ film at 1.4 K under zero magnetic field~\cite{deng2020}. Meanwhile, the axion insulator states characterized by a robust zero plateau of Hall resistance together with large longitudinal resistances are detected in our experiments studying an even-layer (6 SLs) sample~\cite{liu2020nat_mater}. Moreover, a quantum phase transition from axion insulator to Chern insulator is realized experimentally, when the magnetic state is driven from A-type AFM to FM by magnetic fields~\cite{liu2020nat_mater}. In another work, our collaborators  performed transport measurements on thicker films, and observed high-Chern number Chern insulator states in the FM state~\cite{ge2020}. The increase of Chern number with film thickness provides an indirect evidence that the FM MnBi$_2$Te$_4$ is a topological WSM. These experimental data systematically and consistently support our theoretical predictions of magnetic topological physics in MnBi$_2$Te$_4$.  

Overall, the preliminary works have well demonstrated that there exist extremely rich topological physics and outstanding material properties in MnBi$_2$Te$_4$, which is truly a promising material platform to study the interplay of dimension, symmetry, magnetism and topology and can serve as a ``hydrogen model'' system for exploring emergent topological quantum phenomena.

\subsection{Prediction of high-temperature QAH insulator LiFeSe}
The existing intrinsic QAH insulators, including MnBi$_2$Te$_4$ films~\cite{deng2020} and twisted bilayer graphene~\cite{serlin2020}, have working temperatures far below the room temperature, which is limited by the low magnetic transition temperature originated from the weak magnetic coupling. It is thus preferred to utilize stronger exchange coupling mechanisms for obtaining 2D ferromagnets with high Curie temperature $T_C$. In nature, a few transition metals in their bulk are well known to have very high $T_C$. For instance, $T_C$ of Fe, Co, and Ni are 1043, 1400 and 627 K, respectively. Materials including 2D layers of 3$d$ transition metals are thus of particular interest. Experimentally, a state-of-the-art gating technique has been developed to inject a large amount of lithium (Li) ions into layered materials for tuning magnetic properties~\cite{deng2018}. This motivates us to apply the strategy of Li decoration to change magnetic properties of existing vdW layered materials and search for high-temperature QAH insulators.

\begin{figure}
  \includegraphics[width=0.49\textwidth]{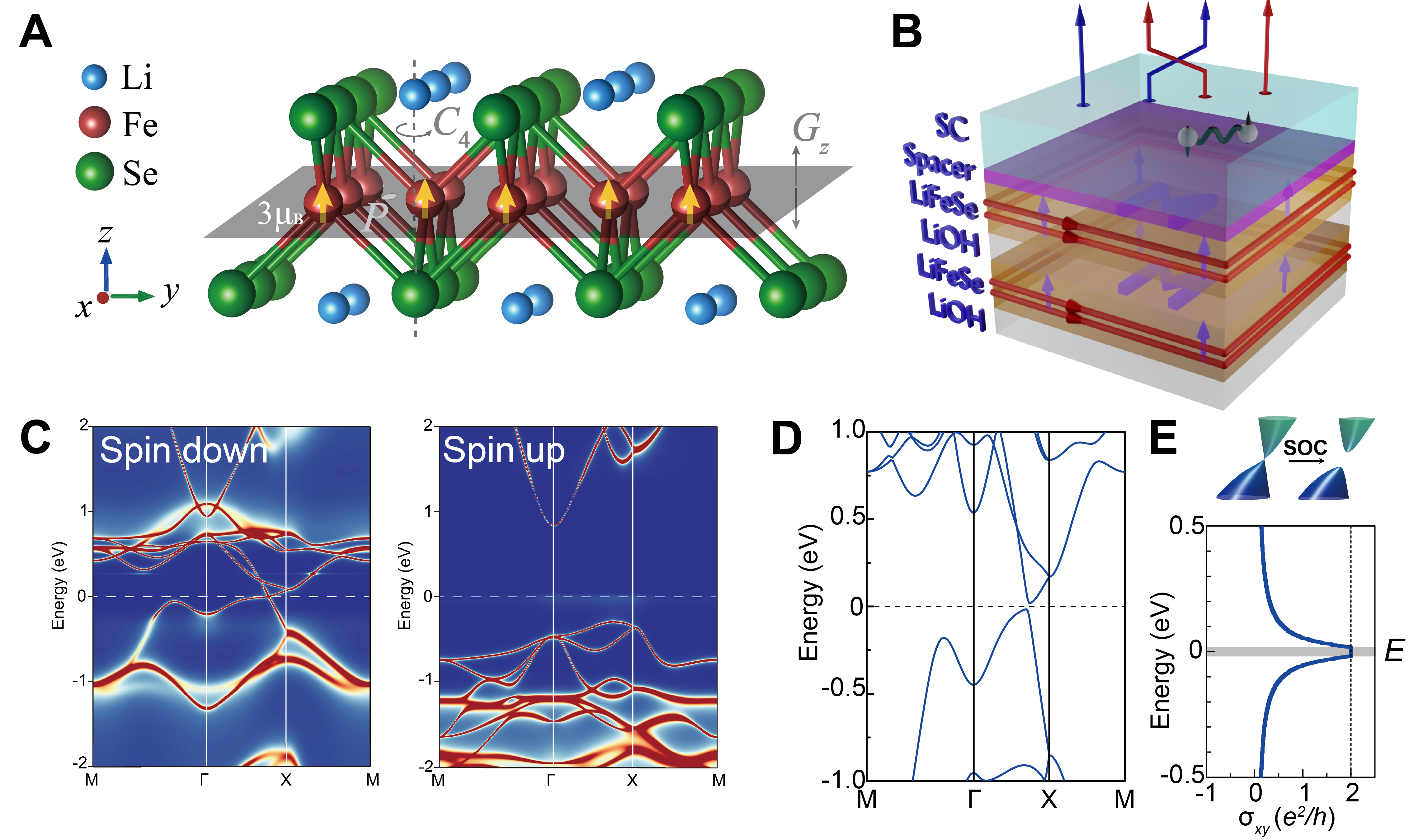}
  \label{fig6}
  \caption{(A) Atomic and magnetic structures of monolayer LiFeSe. (B) A natural heterostructure between superconductor and QAH insulator (LiFeSe) built by gate-controlled Li injection. (C) Band structure of LiFeSe calculated by DFT+DMFT excluding the SOC. (D,E) Topological nontrivial band structure and the anomalous Hall conductance \(\sigma_{xy}\) as a function of Fermi energy calculated by DFT including the SOC. Figure adapted from ~\cite{li2020prl}.}
\end{figure}

Based on first-principles calculations, we predicted that the Li-decorated iron-based superconductor materials, LiFe$X$ ($X=$ S, Se, Te), are ideal candidate materials to realize high-temperature QAH effects~\cite{li2020prl}. The FeSe-family materials are high-temperature superconductors having vdW layered structures. A full adsorption of Li atoms on monolayer FeSe leads to the formation of LiFeSe (Fig. 6A), which is dynamically and thermally stable as supported by phonon calculations and molecular dynamic simulations. Though sharing the same crystal symmetry as FeSe (space group $P4/nmm$), this new 2D material has unexpectedly extraordinary material properties as shown below.

The monolayer LiFeSe is an extremely robust 2D ferromagnet, whose FM exchange coupling strength is stronger than any other known 2D materials, giving $T_C$ far beyond room temperature~\cite{li2020prl}. Due to the small Fe-Fe distance (2.6 $\AA$), the 3$d$ orbitals of neighboring Fe atoms overlap significantly with each other. The electron injection from Li to Fe increases the 3$d$ orbital occupation and thus stabilizes a high-spin state. As a result, the electron hopping from one Fe to the other is allowed (forbidden) by the Pauli exclusion principle if their spins are aligned parallelly (anti-parallelly), leading to strong FM kinetic exchange~\cite{coey2010,eremin1981}. Interestingly, the lattice parameters and magnetic moments of Fe atoms are essentially the same as in monolayer Fe that has been experimentally fabricated in graphene pores~\cite{zhao2014}. Whereas the monolayer Fe is unstable on its own, it can stably exist within the sandwich structure of LiFeSe.

In the absence of SOC, LiFeSe is polarized by the strong ferromagnetism to be a half Dirac semimetal, which displays a large insulating band gap for spin up and a Dirac-like linear band crossing for spin down (Fig. 6C). The two pairs of Dirac points, whose band degeneracies are protected by the $M_x$ and $M_y$ mirror symmetries, appear in the Brillouin zone along $\Gamma$-$Y$ and $\Gamma$-$X$, respectively. The predicted electronic structure is ensured by symmetry and insensitive to computational methods, as confirmed by various advanced approaches including density functional theory (DFT) in combination with dynamical mean-field theory (DMFT), namely DFT+DMFT. The inclusion of SOC in combination with an out-of-plane ferromangetism breaks all the mirror symmetries, opening large Dirac gaps of $\sim$35 meV (Fig. 6D,E). As each Dirac gap contributes a Berry phase of $\pi$ by occupied electrons, the whole system has a Berry phase of $4\pi$ in total, corresponding to $C=2$. The nonzero topological Chern number is further verified by calculations anomalous Hall conductance (Fig. 6E) and edge states. Conclusively, monolayer LiFeSe is a large-gap QAH insulator with room-temperature 2D ferromagnetism, which is promising for realizing high-temperature QAH effects.

Moreover, the 3D LiOH-LiFeSe, possibly obtainable from the existing material LiOH-FeSe by Li injection, is predicted to be a 3D QAH insulator~\cite{jin2018}. This new topological state have macroscopic number of chiral conduction channels, which may find important applications in low-dissipation electronics. By gating-controlled space-modulated Li injection, heterostructures between a high-temperature superconductor and a QAH insulator could be constructed in the FeSe-related materials (Fig. 6B). Such kind of systems, if realized experimentally, could open great opportunities to explore chiral topological superconductivity and Majorana fermion related physics and applications at high temperatures~\cite{qi2010prb,lian2018}.

Finally let us discuss possible experimental evidences. Recent works discovered that the injection of Li into (Li,Fe)OH-FeSe and (Li,Fe)OH-FeS can induce a superconducting to ferromagnetic insulating transition, leading to the emergence of a high-$T_C$ ferromagnetism and strong anomalous Hall effects in the insulating state~\cite{ma2019,lei2019}. The observed effects were tentatively attributed to be caused by the Fe atoms substituted from (Li,Fe)OH. However,  the major physical properties can be reproduced by first-principles calculations excluding the substituted Fe~\cite{li2020prl}, which supports our theoretical prediction. Despite the preliminary works, the striking prediction of high-temperature QAH states in LiFeSe-family materials is awaiting for experimental proof.

\subsection{Different physical scenarios to realize QAH insulators}
As a brief summary, we have predicted four types of novel 3D magnetic topological states in realistic materials, including 3D AFM TI, FM WSM and FM high-order TI in MnBi$_2$Te$_4$-family materials and their heterostructures, as well as 3D QAH insulator in LiOH-LiFeSe. By reducing the physical dimensions from 3D to 2D, quantum confinement of these topological materials all can generate QAH states. This offers four different physical scenarios to realize  QAH insulators with unusual properties of high Chern number, ``pseudo-3D'' and high working temperature. The findings greatly enrich the whole family of QAH materials and facilitate the study of QAH-related physics and applications.

\section{Discovery of novel quantum physics in 2D materials stanene and its derivatives}

The subject of 2D materials has attracted intensive research interests benefiting from their extraordinary quantum effects and great controllability of material properties. While current research mostly focuses on 2D materials exfoliated from vdW layered materials, a much broader class of non-layered 2D materials remain largely unexplored. They are of great interest owing to their the potential to generate diverse material properties with unusual chemical stoichiometries and atomic structures that normally cannot find counterparts from 3D bulk materials~\cite{gou2018}. However, the substrate effects and the interface physics become crucial and must be carefully considered to understand and control material properties, which represents a formidable challenge for the material study. In this context, guidances provided by first-principle material simulation and a close collaboration between theory and experiment are essential to overcome the challenge. 

Here, using stanene and its derivatives as representative non-layered 2D materials, we will first review our theoretical predictions of novel quantum states and effects in the new materials, including large-gap QSH states~\cite{xu2013}, a new 2D semiconductor with unusual defect physics~\cite{gou2018,gou2020}, topological thermoelectric size effects~\cite{xu2014}, and type-II Ising superconductivity~\cite{wang2019prl}. Then we will show that the predicted materials and effects have mostly been realized through close collaborations with experimentalists.

\subsection{Large-gap QSH states in stanene}
TI materials can be classified into two types in 2D and 3D by their physical dimensions, which give gapless topological Dirac fermions on their edges and surfaces, respectively~\cite{hasan2010,qi2011}. Both kinds of topological boundary states are protected against backscattering by TRS. Differently, elastic scatterings at any angle other than 180 degrees are allowed for the surface states of 3D TIs but forbidden for the edge states of 2D TIs, which renders 2D TIs more preferable for low-dissipation electronics. However, in contrast to the extensive studies of 3D TIs, much less research progresses have been achieved on 2D TIs or QSH insulators, despite the fact that the early theoretical and experimental discoveries of TIs were made in 2D material systems of graphene and HgTe/CdTe quantum well~\cite{kane2005,bernevig2006,konig2007}.

Compared to the 3D counterpart, the theory of 2D TIs looks simpler, but the experimental realization is turned out be much more complicated. The key issue is on the materials, which get easily affected by environment and difficult to fabricate and characterize when thinned down to the 2D limit. The proposal of finding QSH effect in graphene is theoretically beautiful but limited by the extremely weak SOC effects of the material, which can be hardly realized by experiment~\cite{kane2005}. The QSH effect was first experimentally discovered in HgTe/CdTe quantum well and further observed in InAs/GaSb quantum well at very low temperatures~\cite{bernevig2006,konig2007,liu2008,knez2011}. These quantum well systems have kept as the only experimentally available 2D TI materials, before the predictive discovery of stanene- and WTe$_2$-related materials~\cite{xu2013,qian2014}. Model material systems with simple structures and large QSH gaps are obviously desired for promoting fundamental research and practical applications.

\begin{figure}
  \includegraphics[width=0.49\textwidth]{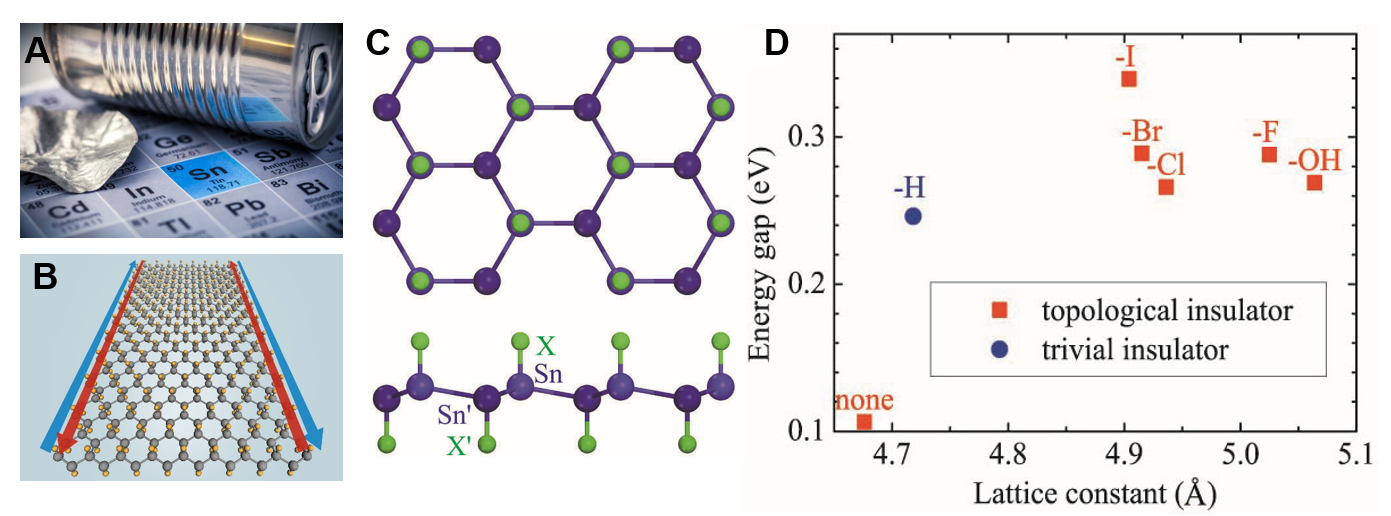}
  \label{fig7}
  \caption{(A) Element Sn in the periodic table. (B) Illustration of QSH edge states in decorated stanene. (C) Atomic structures and (D) band gaps, lattice constants, and topological properties of 2D stanene-related materials Sn\(X\), where \(X=\)-H, -I, -Br, -Cl, -F, OH. Figures (C,D) from ~\cite{xu2013}.}
\end{figure}

Based on first-principles calculations, we predicted that 2D thin films of $\alpha$-Sn are ideal candidate materials of large-gap QSH insulators (Fig. 7)~\cite{xu2013}. The thinnest $\alpha$-Sn (111) film has a buckled honeycomb lattice, which corresponds to the tin (Sn) counterpart of graphene, namely stanene. The ``magic'' honeycomb structure of stanene introduces QSH states near the K/K' valleys by the $p_z$ orbitals, similar as found in graphene~\cite{kane2005,liu2011prb}. Such kind of QSH states, however, could only exist in ideal experimental conditions, because the partially occupied $p_z$ orbitals of stanene are chemically active and easily affected by substrates and adsorbates. Importantly, if fully saturating the $p_z$ orbitals with chemical groups $X$ ($X=$ -F, -Cl, -Br, -I, -OH, etc.), large band gaps are opened at K/K', the material becomes chemically stable, and a topological band inversion between the bonding $s$ orbitals and the anti-bonding $p_{xy}$ orbitals of Sn occurs at the Brillouin zone center ($\Gamma$ point), giving rise to a new kind of QSH states. Caused by the strong SOC effects of Sn $p_{xy}$ orbitals, these QSH states have sizable band gaps $\sim$0.3 eV, suitable for room-temperature uses. Moreover, their properties are highly tunable, for instance, by varying chemical groups or applying lattice strains, which could potentially applied to design topological electronic devices~\cite{molle2017}. Furthermore, we predicted that the large-gap QSH states can survive in stanene samples coupled strongly with substrates (Fig. 8A)~\cite{xu2015prb} or in $\alpha$-Sn films of different crystallographic orientations and film thickness~\cite{li2019spin}, which facilitates the experimental realization.

Our collaboration with experimentalists leads to fruitful progresses on the study of stanene. In 2015, we for the first time fabricated monolayer Sn on Bi$_2$Te$_3$(111) by MBE and identified its atomic and electronic structures by STM and ARPES, respectively, which confirmed the existence of stanene structure~\cite{zhu2015}. However, the stanene sample is metallic on the TI substrate. After long-time intensive experimental efforts and guided by theory, we successfully obtained the first insulating stanene sample on PbTe(111) (Fig. 8D)~\cite{zang2018}. Unfortunately, a careful comparison between experiment and theory indicates that the insulating sample is topologically trivial, due to the small lattice constant constrained by the substrate. One can either enlarge the lattice constant or increase the layer numbers to achieve a topological band inversion. For an ultraflat stanene with a large lattice constant fabricated on Cu(111), our experiments indeed observed a band gap of $\sim$0.3 eV opened by the SOC and the existence of metallic edge states within the gap (Fig. 8B)~\cite{deng2018nat_mater}. Moreover, we experimentally studied the quantum size effects in few-layer stanene and unexpectedly discovered robust superconductivity in the material (Fig. 8E)~\cite{liao2018}. The finding is somewhat surprising, considering that the bulk $\alpha$-Sn is non-superconducting. Owning to the coexistence of topological and superconducting states, $\alpha$-Sn films become promising material candidates to explore topological superconductivity and Majorana fermions. Furthermore, we theoretically predicted a new type of Ising superconductivity in stanene~\cite{wang2019prl} and the novel quantum state was soon found by our experimental collaborators (Fig. 8F)~\cite{falson2020}, as to be discussed below. The interplay of theory and experiment continually gives us surprises in this research direction.

\begin{figure}
  \includegraphics[width=0.49\textwidth]{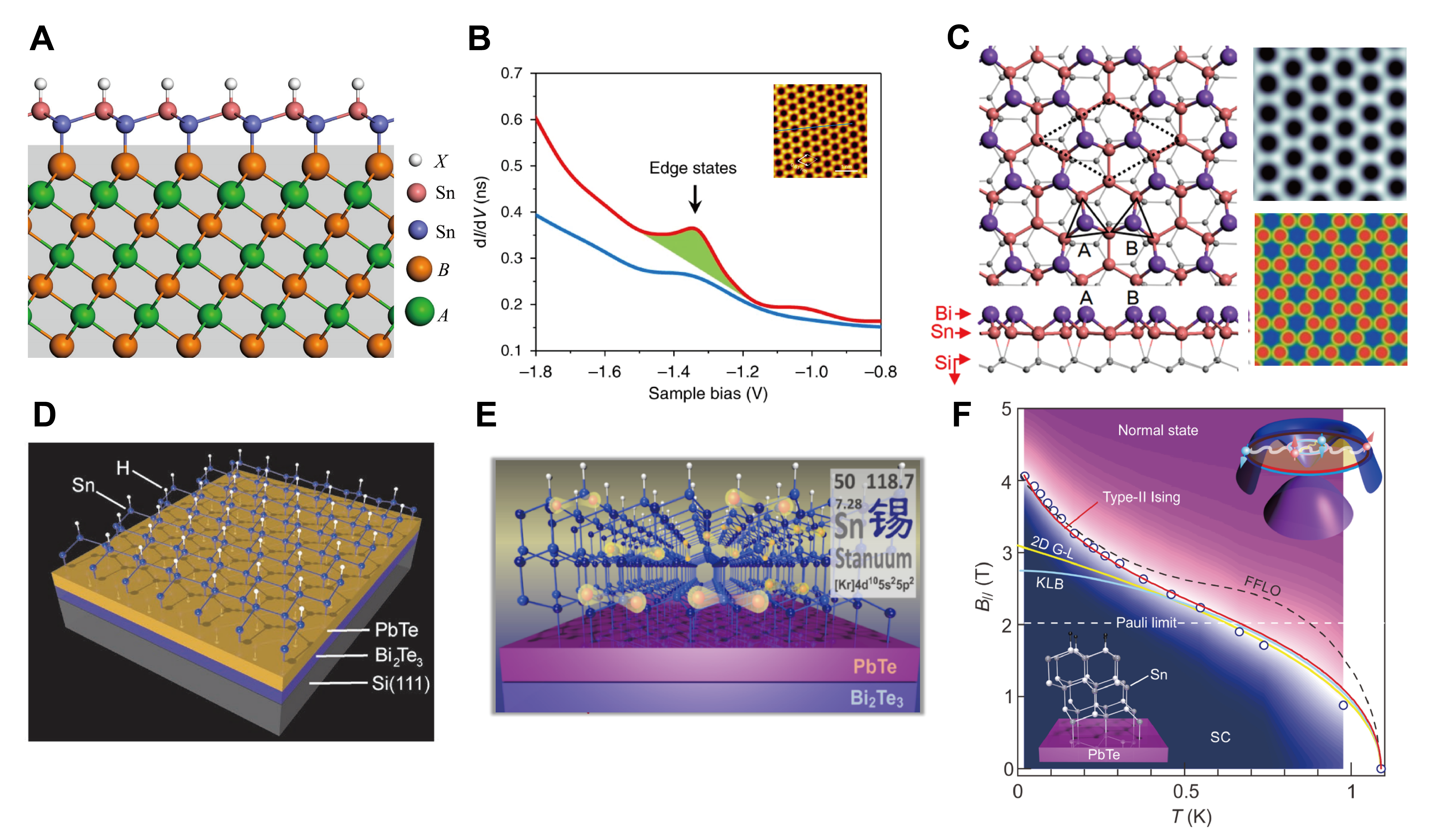}
  \label{fig8}
  \caption{(A) Monolayer Sn\(X\) bonding strongly with the $AB$(111)-$B$ substrate, where $A$ = Pb, Sr, Ba, $B$ = Se, Te, and $X$ = H, F, Cl, Br, I, etc.  (B) STM data of ultraflat stanene on Cu(111). (C) Atomic structure and STM images of 2D semiconductor Sn\(_2\)Bi grown on Si(111). (D-F) schematic diagrams showing an insulating stanene on PbTe~\cite{zang2018} and 2D superconductivity in few-layer stanene grown on PbTe~\cite{liao2018}. (F) The experimental evidence of type-II Ising superconductivity in few-layer stanene~\cite{falson2020}. Figures (A) from ~\cite{xu2015prb}, (B) adapted from ~\cite{deng2018}, and (C) adapted from ~\cite{gou2018}.}
\end{figure}

\subsection{A new 2D semiconductor with unusual defect physics}
A new 2D semiconductor Sn$_2$Bi, which is composed of a unique honeycomb lattice of Bi atoms connected with each other by bonding with a triangular net work of Sn atoms beneath (Fig. 8C), has been discovered on Si(111) by our theoretical and experimental works~\cite{gou2018}. The original motivation is to grow monolayer Sn on Si(111)-$\sqrt3 \times \sqrt3$-Bi. The STM experiments observed a high-quality single-layer honeycomb structure with an insulating gap of $\sim$0.8 eV. By considering thousands possible structures, first-principles calculations identified this stable phase of Sn$_2$Bi, which can well reproduce all the experimental results of STM and ARPES, thus conclusively confirmed the predicted phase.

The non-layered 2D material Sn$_2$Bi has unusual chemical stoichiometry and atomic structure, which cannot find a counterpart in bulk materials~\cite{gou2018}. The electronic structure is also quite special, showing strong SOC effects and the coexistence of dispersive valence bands and flat conduction bands. The high electron-hole asymmetry is caused by the unique atomic structure. Specifically, the Bi atoms do not bond directly with each other but indirectly via Sn atoms. The Bi-Sn orbital hybridization is strong in valence bands, leading to nearly free hole excitations; whereas the orbital hybridization becomes weak in conductance bands, giving rise to strongly localized electron excitations. The special electronic properties might be useful for exploring strong-correlation physics and device applications.

In an unpublished work, we found unusual defect physics in Sn$_2$Bi that can realize quinary charge states in solitary defects~\cite{gou2020}. The realization of high-charge in-gap defect states in semiconductors is demanded by charge-based quantum devices but is a considerably challenging task. This is limited by the fact that the defect Coulomb charing energy increases with the number of defect charges, which could push the defect levels outside the band gap and thus forbid high-charge defect states. One might suggest to use larger band gaps to keep charged defect states in-gap. However, materials with larger insulating gaps usually have more localized defect charge distributions and thus stronger Coulomb repulsions. Therefore, the pursuit is challenged by the conflicting requirements of delocalization for low charging energy and localization implied by large band gap. A compromise can be achieved in Sn$_2$Bi, where the band gap is sizable and defect charge states are unusually delocalized, benefiting from the network structure composed of ``metallic'' elements with similar electronegativities. Thus, an ultralow defect Coulomb charing energy is realized in a large-gap system, making high-charge in-gap defect states feasible. The finding suggests a new route to realize high-charge defect states for developing charge-based quantum devices.

\subsection{Topological thermoelectric size effects}
The discovery of TIs sheds new lights on the research of thermoelectrics, since many TIs, like (Bi,Sb)$_2$Te$_3$, actually are excellent thermoelectric (TE) materials~\cite{xu2014,xu2016,xu2017,he2017,mao2018,fu2020apl_mater}. In fact, TI and TE materials share similar material traits of heavy elements and narrow band gaps~\cite{xu2016,xu2017}. So that TIs can have strong SOC effects to induce topological band inversions, whereas TE materials can have low lattice thermal conductivities as well as large power factors. How to utilize topological states to improve TE performance, however, is an important open question.

\begin{figure}
  \includegraphics[width=0.49\textwidth]{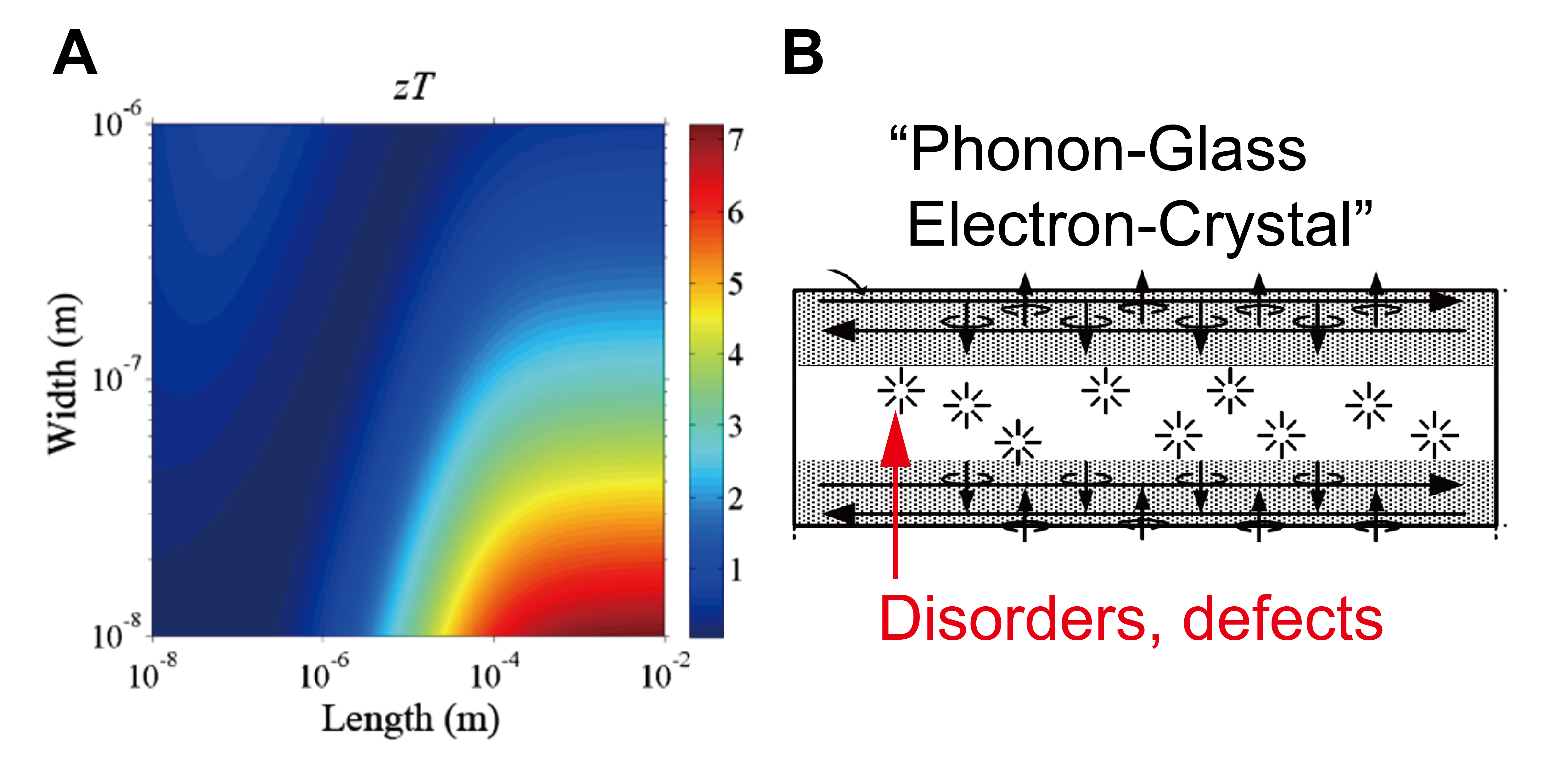}
  \label{fig9}
  \caption{(A) The geometric size dependence of thermoelectric figure of merit $zT$ for the 2D TI fluorinated stanene at 300 K. (B) The realization of ``phonon-glass electron-crystal'' in QSH insulators. Figure (A) from ~\cite{xu2014}.}
\end{figure}

We predicted that TE effects become strongly size dependent in TIs, and the optimization of geometric sizes could significantly enhance the TE performance (Fig. 9)~\cite{xu2014}. Using stanene as an example, we showed that the TE performance is greatly improved by optimizing the contribution of edge sates. Specifically, by introducing defect and disorders in the material and by reducing (increasing) the materials width (length), the phonon contribution get depressed, whereas the edge-state contribution is almost unaffected. This effectively decouples contributions of electrons and phonons, thus realizing the concept of ``phonon-glass electron-crystal'' (Fig. 9B). The way of improving TE performance by geometric optimization is efficient and easy to control, which generally applies to topological materials in 2D, including QSH and QAH insulators~\cite{he2017,mao2018,fu2020apl_mater}.

\subsection{Type-II Ising superconductivity}

Ising superconductors belong to a novel kind of superconductors that are protected against external magnetic fields by the SOC effects and can display an upper critical field $B_{c2}$ exceeding the classical Pauli limit $B_p = 1.86 T_{C}$~\cite{clogston1962,chandrasekhar1962} ($B_p$ and the zero-field critical temperature $T_{C}$ are in units of Tesla and Kelvin, resepectively)~\cite{lu2015,saito2016,zhou2016prb}, showing promise for potential applications in high-speed maglev, high-field magnet, high-energy collider, etc. In the traditional physical scenario of Ising superconductivity, the inversion symmetry must be broken and an out-of-plane mirror reflection symmetry is required. Thus few candidate materials have been found, which are limited to the monolayers of transition metal dichalcogenides in the 2H phase~\cite{lu2015,saito2016,zhou2016prb}.

A common knowledge is that the SOC-induced spin splitting is zero if both inversion symmetry and TRS are preserved, implying zero SOC fields. Guided by this analysis, previous works on Ising superconductivity only considered inversion asymmetric materials. We found an important loophole in the symmetry analysis that is applicable to single-orbital systems only. Realistic materials usually have orbital degeneracies protected by crystalline symmetries, which should be described by the physics of multiple orbitals.

\begin{figure}
  \includegraphics[width=0.49\textwidth]{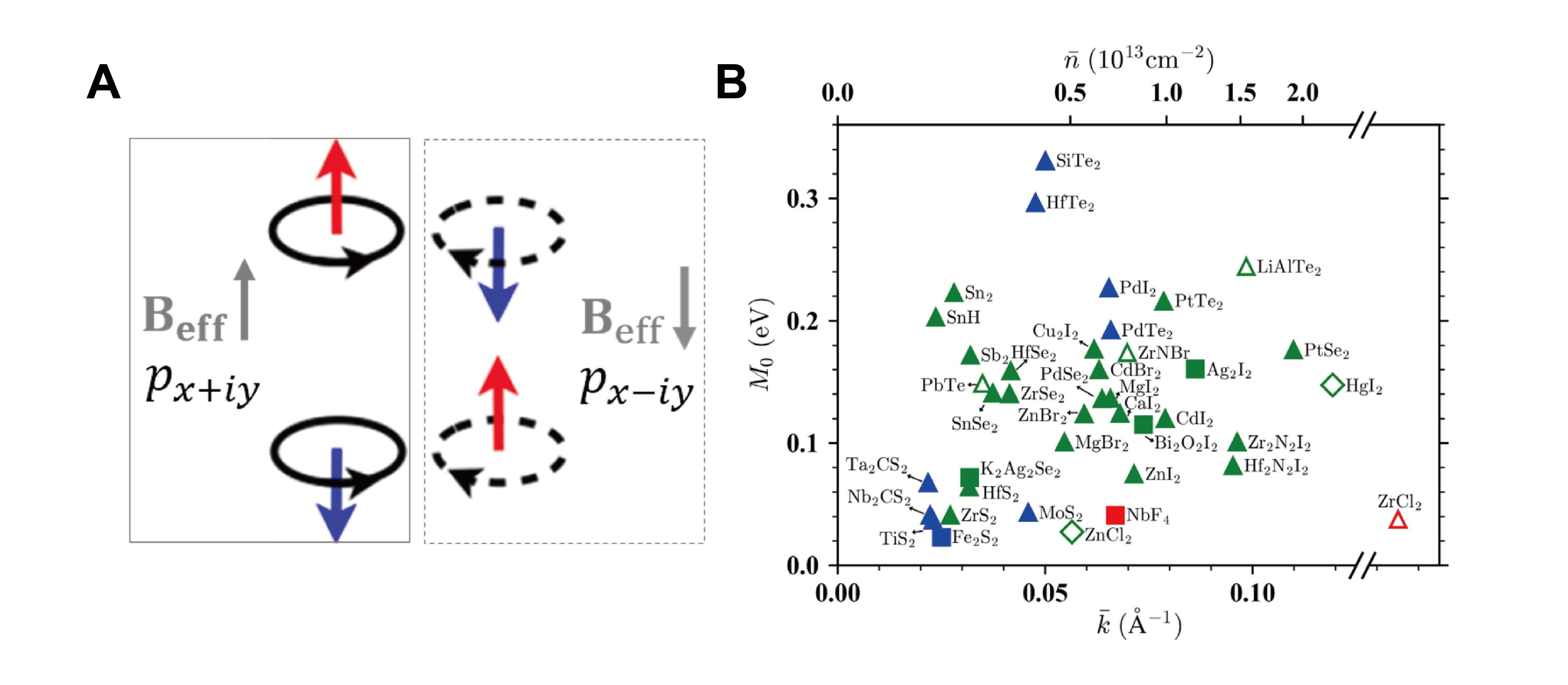}
  \label{fig10}
  \caption{(A) The spin-orbital locking giving rise to opposite Zeeman-like fields {\bf{B}}$_{\rm{eff}}$ for opposing orbitals $p_{x\pm iy}$ .  (B) Candidate materials to study type-II Ising superconductivity.  Figure (B) from ~\cite{wang2019prl}.}
\end{figure}

Based on the above observation, we predicted a new type of Ising supercondcuting pairing mechanism that does not require inversion symmetry breaking, called type-II Ising superconductivity~\cite{wang2019prl}. The key physics can be well demonstrated in stanene that is inversion symmetric and has degenerate $p_{xy}$ orbitals, as ensured by the $C_{3\rm{v}}$ symmetry. The $p_{xy}$-related Bloch states have an four-fold degeneracy at $\Gamma$ when excluding the SOC. They are split into two spin-degenerate copies by the SOC, including the $|j_z = \pm 3/2\rangle$ (i.e., $|p_{x+iy},\uparrow\rangle$ and $|p_{x-iy},\downarrow\rangle$) and $|j_z = \pm 1/2\rangle$ (i.e., $|p_{x+iy},\downarrow\rangle$ and $|p_{x-iy},\uparrow\rangle$) states. Importantly, the SOC results in a  spin-orbital locking, which gives opposite Zeeman-like fields for opposing orbitals $p_{x\pm iy}$ (Fig. 10A). As the circular orbital motions are in-plane, the generated Zeeman-like fields are along the out-of-plane direction. Such kind of effective SOC-fields are extremely large ($\sim$10$^3$ Tesla in stanene), which greatly suppresses the Zeeman splitting caused by external in-plane magnetic fields, leading to an enhanced $B_{c2}$ significantly beyond $B_p$. By high-throughput first-principles calculations, we predicted many candidate materials for experimental studies (Fig. 10B). This work greatly enriches the research on new physics and materials of Ising superconductivity.

As recent experimental progresses, our collaborators studied the influence of magnetic fields on superconductivity in few-layer stanene, and observed that the in-plane upper critical field indeed exceeds the Pauli limit, which provides a smoking-gun evidence for the existence of type-II Ising superconductivity~\cite{falson2020}. The theoretical prediction has also been experimentally confirmed in other material systems~\cite{liu2020nano_lett}, demonstrating the generality of Ising superconductivity physics.

\section{Conclusion and outlook}
We have demonstrated that the interplay of SOC with magnetism, topology, and superconductivity provides a fertile field to explore novel physics and materials, and the first-principles material prediction guided by fundamental theory and in close collaboration with experiment becomes a powerful paradigm to study quantum materials. As an outlook, an unprecedentedly new paradigm of material research would emerge in the near future by building material databases and using big-data methods. Traditionally, people try to first understand the underlying physical mechanisms by studying specific materials and then predict useful new materials based on the established understanding. Such an order could be reversed. The use of material databases and  big-data approaches is able to predict materials with excellent properties even with little prior physical knowledge. Then mysterious physics could be discovered by investigating the proposed materials. This new research paradigm might significantly change our future scientific research.

\section*{Acknowledgement}
This work is supported by the Basic Science Center Project of NSFC (Grant No. 51788104), the Ministry of Science and Technology of China (Grants No. 2018YFA0307100, and No. 2018YFA0305603), and the National Natural Science Foundation of China (Grant No. 11874035).

\end{document}